\documentclass{article}

\usepackage{arxiv}

\usepackage[utf8]{inputenc} 
\usepackage[T1]{fontenc}    
\usepackage{hyperref}       
\usepackage{url}            
\usepackage{booktabs}       
\usepackage{amsfonts}       
\usepackage{nicefrac}       
\usepackage{microtype}      
\usepackage{lipsum}
\usepackage{graphicx}
\usepackage{caption}
\usepackage{subcaption}
\usepackage{natbib}

\setcitestyle{numbers,open={[},close={]}}

\usepackage{doi}

\usepackage{multicol}

\graphicspath{ {./images/} }

\title{An information retrieval and extraction tool for covid-19 related papers}

\author{ \href{https://orcid.org/0000-0002-1553-8837}{\includegraphics[scale=0.06]{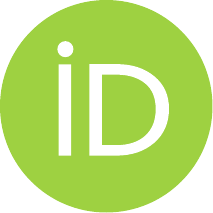}\hspace{1mm}Marcos V.~L.~Pivetta} \\
	Universidade Federal do Rio Grande do Sul (UFRGS) \\
	Rio Grande do Sul, Brazil \\
	\texttt{marcos.pivetta@inf.ufrgs.br} \\
}

\hypersetup{
pdftitle={An information retrieval and extraction tool for covid-19 related papers},
pdfsubject={cs.CL},
pdfauthor={Marcos V.~L.~Pivetta},
pdfkeywords={Health Information Management; Information Storage and Retrieval; Healthcare; COVID-19; Unified Medical Language System},
}

\begin{document}

\maketitle

\begin{multicols}{2}

\begin{abstract}
\textbf{Background:} The COVID-19 pandemic has caused severe impacts on health systems worldwide. Its critical nature and the increased interest of individuals and organizations to develop countermeasures to the problem has led to a surge of new studies in scientific journals. \textbf{Objetive:} We sought to develop a tool that incorporates, in a novel way, aspects of Information Retrieval (IR) and Extraction (IE) applied to the COVID-19 Open Research Dataset (CORD-19). The main focus of this paper is to provide researchers with a better search tool for COVID-19 related papers, helping them find reference papers and hightlight relevant entities in text. \textbf{Method:} We applied Latent Dirichlet Allocation (LDA) to model, based on research aspects, the topics of all English abstracts in CORD-19. Relevant named entities of each abstract were extracted and linked to the corresponding UMLS concept. Regular expressions and the K-Nearest Neighbors algorithm were used to rank relevant papers. \textbf{Results:} Our tool has shown the potential to assist researchers by automating a topic-based search of CORD-19 papers. Nonetheless, we identified that more fine-tuned topic modeling parameters and increased accuracy of the research aspect classifier model could lead to a more accurate and reliable tool. \textbf{Conclusion:} We emphasize the need of new automated tools to help researchers find relevant COVID-19 documents, in addition to automatically extracting useful information contained in them. Our work suggests that combining different algorithms and models could lead to new ways of browsing COVID-19 paper data.
\end{abstract}

\keywords{Health Information Management \and Information Storage \and Retrieval \and Healthcare \and COVID-19 \and Unified Medical Language System}

\section{Introduction}
The disease caused by the new coronavirus, named COVID-19 (Coronavirus Disease 2019), has led to a worldwide public health crisis. The World Health Organization (WHO), in its strategic plan to combat COVID-19, highlights the urgent need for \textbf{research, innovation, and knowledge sharing} about the issue \cite{WorldHealthOrganization2020COVID19UPDATE}. Because of the rapidly increasing number of scientific papers about the topic, scientists report not having time to identify the core claims and values added of every COVID-19 related paper that is being published \cite{Brainard2020ScientistsAfloat}. 

Automated tools to help researchers of COVID-19 find relevant documents and information within them are a growing necessity and could solve the previously described hindrances. We introduce a tool, named CORD-19 Knowledge Tool (CORD-19 KTool) \footnote{\url{https://github.com/pivettamarcos/CORD-19_KTool}}, that can help researchers in this task by means of (i) recommending CORD-19 papers based on LDA topic modeling combined with research aspect classification and (ii) extracting named entities and linking them to the respective UMLS concept.

In the following introductory sections, we give the reader an insight into the existing sources of curated COVID-19 paper data. We also describe the current situation of COVID-19 paper recommendation systems and the usefulness of Named Entity Recognition (NER) systems for COVID-19 studies. These sections also provide contextualization to our own approach to both of these tasks. 

\subsection{Data sources}

Initiatives have been taken to create a reliable source of information about COVID-19 related studies. Resources such as the  COVID-19 Open Research Dataset (CORD-19) \cite{Wang2020CORD-19:Dataset} and LitCovid \cite{Chen2020KeepResearchb} aim to gather an ever-increasing number of papers about COVID-19 and related coronavirus diseases for data mining purposes.

Released in March 2020, CORD-19 is one of the main curated sources of articles about COVID-19 and related coronaviruses. The data set was created from a collaboration between the Allen Institute for AI (AI2) and organizations such as The White House Office of Science and Technology Policy (OSTP) and National Library of Medicine (NLM).

CORD-19 contains over 1 million articles from its last update in June 2022. It is structured in a way that facilitates the access by automated tools. To achieve that, it adopts some guiding principles:
\begin{itemize}
    \item standardization of paper's metadata
    \item elimination of duplicate papers
    \item machine readability 
\end{itemize}

The broader topics presented in CORD-19 provide a diverse but complex source for text mining purposes.

\begin{figure*}
  \centering
  \includegraphics[width=0.8\textwidth]{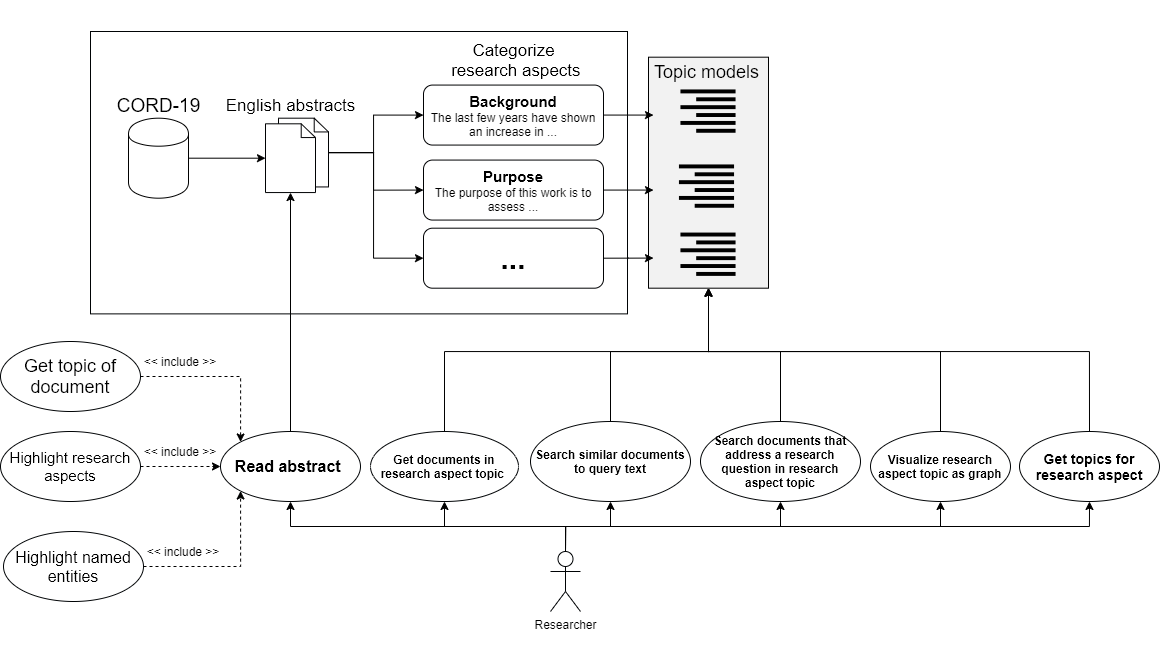} 
  \caption{Diagram describing the tool's use cases and interaction with CORD-19 abstracts and the topic models generated with LDA}
  \label{fig:diagram}
\end{figure*}

\subsection{Paper recommendation}

Many different studies have presented Information Retrieval (IR) systems for COVID-19 paper recommendation. All these systems tackle the problem in a different way. For example, \cite{Sonbhadra2020TargetApproach} combined machine learning algorithms such as K-means, DBSCAN, and HAC to group COVID-19 papers based on the predefined research question they answer. \cite{Esteva2020CO-Search:Summarization} use a combination of Siamese-BERT model and two keyword models to better score retrieved documents in the context of COVID-19 research. 

The application of Latent Dirichlet Allocation (LDA) \cite{Blei2003LatentAllocation} to model the topics of COVID-19 papers has also shown positive results for retrieval tasks. LDA can help researchers organize textual information by modeling the relevant topics from a set of documents. An LDA model generates topics inherent to the documents, using a probability distribution that ensures that all topics obey a Dirichlet polynomial prior distribution \cite{Wu2014TopicResearch}. The works of \cite{Verspoor2020COVID-SEE:Research}, \cite{Wise2020COVID-19Literature} and \cite{Herlihy2020AutomatedModeling} describe systems that provide document exploration based on salient topics in COVID-19 studies and social media posts. 

Topic probability distributions from LDA can be used to aid document search. Determining the most similar documents to a document based on a query text can be done programmatically. Algorithms such as K-Nearest Neighbors (KNN) can classify this new query text based on the class of the nearest previously classified objects \cite{Cover1967NearestClassification}, using topic probability as the feature. 

We could go more in depth with the LDA strategy by combining it to a particular text segment. This local topic modeling strategy increases search granularity and has shown measurable benefits, like presented in the work of \cite{Arnold2019SECTOR:Classification}. To achieve this, we need to classify the sentences or paragraphs in the paper and assign them a role. After that, we can provide topic modeling results to each local text segment, e.g., the introduction, methods or results.

\subsubsection{CODA-19}

The COVID-19 Open Research Dataset (CODA-19) \cite{Huang2020CODA-19:Crowd} is a data set constructed by crowd workers that annotated useful textual structure from 10,966 CORD-19 abstracts. These units, or research aspects, include: background, purpose, method, finding/contribution, and other. The research aspects reflect the role of a particular sentence based on the author's intention. The workers followed a strict annotation scheme that was adapted from the works of \cite{Chan2018SOLVENT:Papers} and \cite{Huang2017DISA:Analysis}. 

A great number of studies have shown the benefits of structured abstracts since their introduction in medical research journals \cite{Hartley2004CurrentAbstracts}. Although well received by researchers, many abstracts still lack defined structural units. Automatic annotation of research structure in abstracts has been the subject of many previous works \cite{Dasigi2017ExperimentNetworks, Banerjee2020SegmentingData, deWaard2012VerbEvidence} that also suggest its usefulness in information retrieval tasks. 

The authors of CODA-19 have publicly released a classifier model that uses SciBERT. SciBERT is a language model trained on 1.14 million biomedical texts from Semantic Scholar and is based on the Bidirectional Encoder Representations from Transformers (BERT) \cite{Devlin2018BERT:Understanding} architecture. Bidirectional networks are state-of the art in several NLP tasks, which involve, for example, textual understanding and question answering \cite{Devlin2018BERT:Understanding}.

Based on experiments made by \cite{Huang2020CODA-19:Crowd}, CODA-19 classifier's achieves an overall accuracy score of 0.774. It's precision, recall and F-measure for each research aspect can be seen on Table \ref{coda_performance}.

\begin{tabular}[t]{ |p{3cm}||p{1cm}|p{1cm}|p{1cm}|  }
 \hline
 \textbf{Research aspect} & \textit{P} & \textit{R} & \textit{F1}\\
 \hline
 Background   & .733 & .768 & .750\\
 Purpose &   .616  & .636 & .626\\
 Method & .715 & .636 & .673\\
 Finding & .783 & .775 & .779\\
 Other & .794  & .852 & .822\\
 \hline
\end{tabular}
\captionof{table}{Performance of the pre-trained Sci-Bert model, as described by Huang et al. \, cite{Huang2020CODA-19:Crowd} \\}
\label{coda_performance}

We use LDA topic modeling to identify latent topics for each individual research aspect identified by CODA-19 in CORD-19 abstracts.

\subsection{Named Entity Recognition (NER)}

Named Entity Recognition (NER), a subtask of Information Extraction (IE) \cite{Nadeau2007AClassification}, aims to automatically extract named entities (names of organizations, locations, quantities, dates, etc.) from text. Scientific papers about COVID-19 bring about many different words from the fields of virology and biomedicine that are important to the understanding of the text, such as the name of: viruses, chemical compounds, diseases, viral components, genes, among others. These words must be extracted with some computational method, such as a NER system. 

\subsubsection{Unified Medical Language System (UMLS).} Works from \cite{Michel2020KnowledgeResearch, Verspoor2020COVID-SEE:Researchb} show that linking named entities to a controlled vocabulary improves the usefulness of IE and IR tasks. A controlled vocabulary system, like the Unified Medical Language System (UMLS) \cite{Bodenreider2004TheTerminology} provides a standardized means of identifying biomedical concepts and their relation to other concepts. We can use this knowledge to improve the IE aspect of our tool.

\subsection{Aim of the study}

In this paper, we describe a researcher-driven tool to aid the process of new hypothesis proposal and the search of reference articles by the researcher. The tool feeds on data provided by CORD-19 and can identify the topics of the abstracts from the data set to aid the search of relevant papers. To achieve more detailed results, we not only identify topics for the whole text, but for each research aspect (introduction, purpose, method and finding/contribution) of the study, using the COVID-19 Research Aspect Dataset (CODA-19) \cite{Huang2020CODA-19:Crowd} pre-trained classifier model. The tool also gives researchers a readily way to visualize named entities and their descriptions according to the UMLS Metathesaurus.

\section{Methods}

\subsection{Tool definition and prototype creation}
\label{sec:headings}

CORD-19 KTool provides the use cases shown in Figure \ref{fig:diagram}. The diagram shows the extraction of COVID-19 abstracts from CORD-19, their segmentation into research aspects, and the respective training of topic models for each segment. It also shows all possible actions the researcher can do.

\subsection{Algorithms}

\subsubsection{Extracting abstracts from CORD-19} 

The abstract of every publication was extracted from the CORD-19's CSV metadata file (version \textit{2020-10-01}), and served as input to the algorithms described in the next sections. Only the abstracts were extracted. This is purposefully done to fit the research aspect classification model, which was trained only on abstracts. Title and publication date were also extracted from the metadata file.

\subsubsection{Preprocessing}
Prior to the application of the Natural Language Processing (NLP) algorithms, The CORD-19 documents went through 4 preprocessing steps:
\begin{enumerate}
    \item \textbf{Only keep abstracts in the English language.} Since the integrated NLP tools used were trained on English texts, only English papers are available in the tool.
    \item \textbf{Segment the abstracts into sentences.} We segmented the abstract into sentences using the en\_core\_sci\_lg SciSpacy model \footnote{https://allenai.github.io/scispacy/}.
    \item \textbf{If the abstract is absent, use the title of the paper as the abstract}
    \item \textbf{Remove stop words.} Words such as ``method", ``introduction", ``conclusion" and other English stop words were removed as they provide no useful information for the topic modeling process.
\end{enumerate}

\subsubsection{Categorizing research aspects}

We used the CODA-19's baseline SciBERT \cite{Huang2020CODA-19:Crowd} pretrained model \footnote{\url{https://github.com/windx0303/CODA-19/tree/master/classification}} to categorize the content of every abstract based on research aspects. With this classifier model, we could assign each sentence in the abstracts a role: background, purpose, method, finding/contribution or other. The SciBERT model was used as it outperformed other models, as presented in the CODA-19 paper. 


\subsubsection{Modeling topics}

For this task, we used the \textit{ktrain} \cite{Maiya2020Ktrain:Learning} wrapper for the Python 3.9.0 programming language. The wrapper offers a readily available set of methods to deal with LDA topic modeling. Separate LDA models were trained for each research aspect of all abstracts. We also trained LDA models that comprise the entirety of all abstracts. Papers that include ``COVID-19'' or ``COVID'' in the abstract were trained to separate models to discriminate them from papers related to other coronaviruses.

The statistical model resultant of LDA can give us an idea of the underlying topics of a collection of documents, based on research aspects and the abstract as a whole. It is also used as a way to cluster documents that cover the same (or similar) topic in an unsupervised way.

We linked the topic modeling process to the research aspects with the purpose of making the search functionality of our tool more fine-grained, as the researcher can focus their search attempt on one of the research aspects or the whole abstract, like previously described.

The \textit{ktrain} parameters used for all topic models are:
\begin{enumerate}

    \item\textbf{Minimum document frequency (min\_df) = 0.001} \\
    \item \textbf{Maximum document frequency (max\_df) = 0.65}\\
    \item \textbf{Number of maximum LDA iterations (lda\_max\_iter) = 15}\\
    \item \textbf{Number of features = 1000000}\\
    
\end{enumerate}

The maximum number of topics for each model (whole text and for each research aspect) is then defined using the default heuristic formula \ref{eq:num} set by \textit{ktrain}:
\begin{equation}
\label{eq:num}
\min{(400, \sqrt{\frac{\textrm{Number of documents in research aspect}}{2}})}	
\end{equation}

The topic modeling done by LDA can yield a lower number of topics than that assigned, meaning some models will differ from the formula.


\subsubsection{Finding similar documents to a query using K-Nearest Neighbors (KNN)}
Using ktrain, we were able to train a recommender model that can find similar documents to a user query, based on the topic probability distribution of the text provided by the user. The ktrain wrapper uses KNN for this purpose. Default parameters were used.

To find documents similar to a search query, we had to:

\begin{enumerate}
    \item \textbf{Train the recommender model }
    
    In the training phase, previously classified sample points are stored as multidimensional vectors in a vector space. In our tool, the topic probability distribution is used as the feature that constructs the vectors. This vector space is then saved as a model for later use.
    
    \item \textbf{Assess the similarity to CORD-19 documents}
    
    To determine the most similar documents to the query, we first determine the topic probability distribution of the user provided text. We do this by using the appropriate topic model and predicting the topic probabilities of the query text. Secondly, we find the k-nearest neighbors to that text using the Euclidean distance metric. The k-nearest neighbors will be the output of this step, meaning they are the most similar documents to that which the user provided.
    
\end{enumerate}


\subsubsection{Searching papers with query string.}
Regular expressions were used to search for documents based on a predefined query of words or sentences.
After conversion of the query string to lower case, the expression \ref{eq:regex} was used for this purpose.
\begin{equation}
\label{eq:regex}
(\verb/r"\b%s\b" & query/)
\end{equation}

The predefined query terms were adapted from the key scientific questions posed on the \textit{kaggle} CORD-19 challenge \cite{Kaggle2020COVID-19CORD-19} in the form of a ``What is known about [topic]?'' question.

\subsubsection{Graphical representation of a topic model.}
The \textit{ktrain} library was used to generate a 2D graphical representation of the topic models. It uses a T-distributed Stochastic Neighbor Embedding (t-SNE) to the topic distribution vector of each paper, reducing it to two dimensions. This way, each paper can be plotted on a 2D graph representing how close each data point (paper) is to another in the topic space.

\subsubsection{Named Entity Recognition and UMLS linking.}

Named Entity Recognition (NER) was achieved using the SciSpacy \cite{Neumann2019ScispaCy:Processing} NER model trained on the CRAFT corpus. CRAFT's annotation can identify terms from nine important biomedical conceptual systems \cite{Bada2012ConceptCorpus}:

\begin{itemize}
    \item Cell Type Ontology (CL) \cite{Bard2005AnTypes.} 
    \item chemical Entities of Biological Interest ontology (ChEBI) \cite{Degtyarenko2008ChEBI:Interest}
    \item NCBI Taxonomy (NCBITaxon) \cite{Sayers2009DatabaseInformation}
    \item Protein Ontology (PRO) \cite{Natale2011TheComplexes}
    \item Sequence Ontology (SO) \cite{Eilbeck2005TheAnnotations.}
    \item Entrez Gene database (EG) \cite{Maglott2011EntrezNCBI}
    \item Gene Ontology (GO) \cite{Ashburner2000GeneBiology}
\end{itemize}

The NER model has an F1 of 75.02. The step of linking the named entity to UMLS \cite{Bodenreider2004TheTerminology} concepts was done using the Entity Linker in SciSpacy. 

\section{Results}
\subsection{Research aspect classification and topic modeling}

CORD-19 1 October 2020 update has a total of 268,170 English abstracts, 72,920 of which mention the words ``COVID-19'' or ``COVID''. We present the number of abstracts that contain each research aspect in Table \ref{num_aspects}. \\

\begin{tabular}[H]{|p{3cm}|p{2cm}| p{2cm}|}
 \hline
  & \textit{All papers} & \textit{Only COVID-19 papers} \\
 \hline
 Background   & 166,160 & 63,243 \\
  
 Purpose   & 94,896 & 37,653 \\
  
 Method   & 76,590 & 27,092 \\
  
 Finding/Contribution   & 175,780 & 65,065 \\
  
 Whole text   & 268,170 & 72,920 \\
 \hline
\end{tabular}
\captionof{table}{Number of abstracts that contain the respective research aspect \\}
\label{num_aspects}

The topic count for every research aspect and the whole text is shown in Table \ref{num_topics}. Topic count is based on the number of unique words present in the available documents for training. \\

\begin{tabular}[H]{|p{3cm}|p{2cm}| p{2cm}|}
 \hline
  & \textit{All papers} & \textit{COVID-19 only papers} \\
 \hline
 Background   & 146 & 96 \\
  
 Purpose   & 130 & 83 \\
  
 Method   & 103 & 84 \\
  
 Finding/Contribution   & 152 & 103 \\
  
 Whole text   & 176 & 101 \\
 \hline
\end{tabular}
\captionof{table}{Topic count for each research aspect \\}
\label{num_topics}

\subsection{List of topics}
\begin{figure*}
\centering
  \centering
  \includegraphics[width=.75\linewidth]{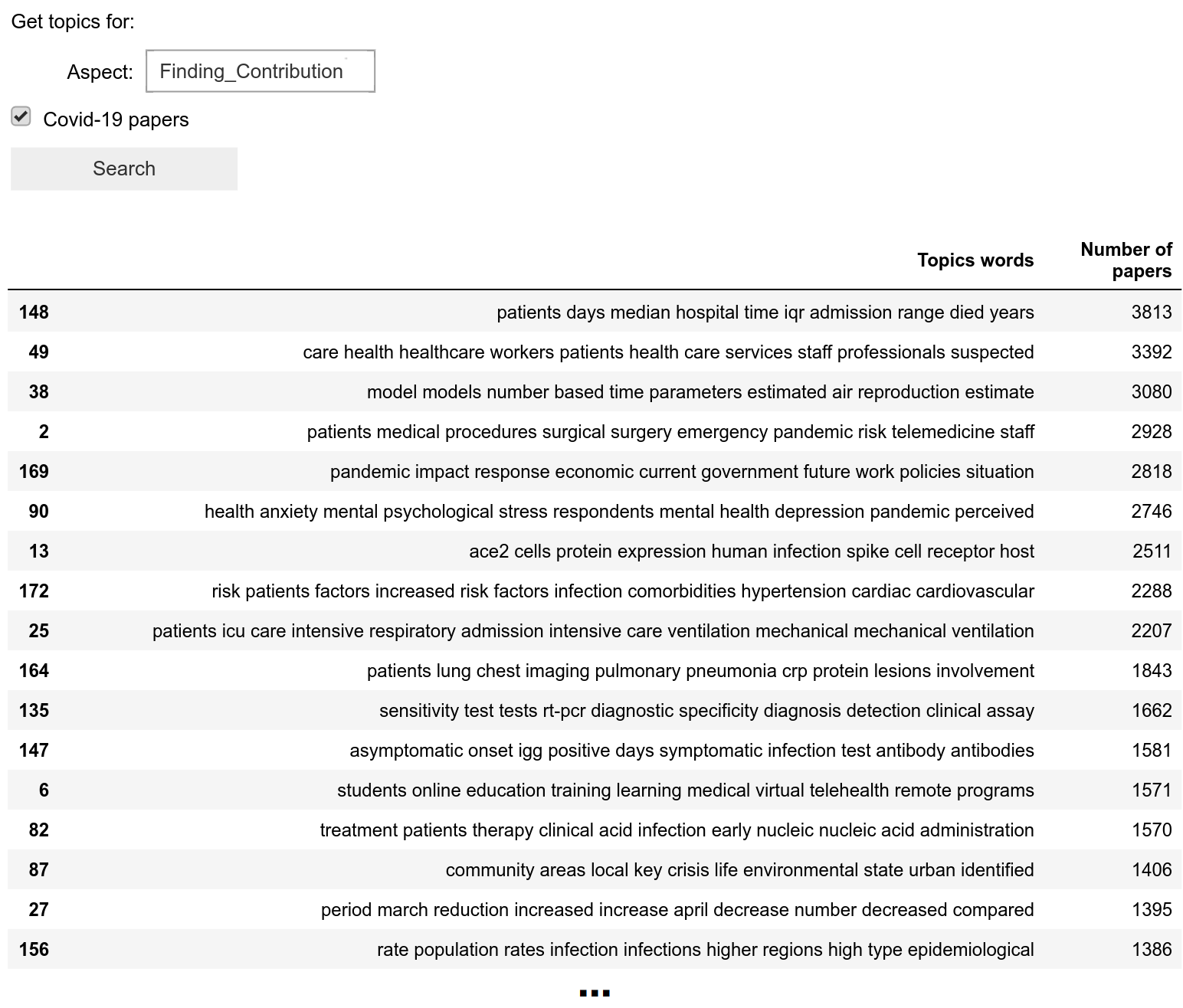}
  \captionof{figure}{List of topics addressed in the finding/contribution section of COVID-19 abstract}
\label{fig:topics}
\end{figure*}

This listing gives the researcher the topics LDA identified for the selected research aspect. The list is ordered by the number of documents in CORD-19 that fall into the respective topic. The list can comprise all of CORD-19 abstracts or be narrowed down to only those related to COVID-19. 
Figure \ref{fig:topics} shows a case study for this functionality. The researcher is looking for the main topics contained in the finding/contribution segment of abstracts about COVID-19. The overall idea of each topic can be deduced from the list of words. For example, one might suggest that topic 148, represented by the words: ``patients, days, median, hospital, time, admission'',  indicates findings related to patient admission to hospital, time spent in hospital, etc. Topic 169, made up of the words ``pandemic, impact, response, economic, current, government, future, work, policies and situation'' could indicate findings related to the pandemic's economic impact, government response to the pandemic, etc.

\begin{figure*}
  \centering
  \includegraphics[width=.75\linewidth]{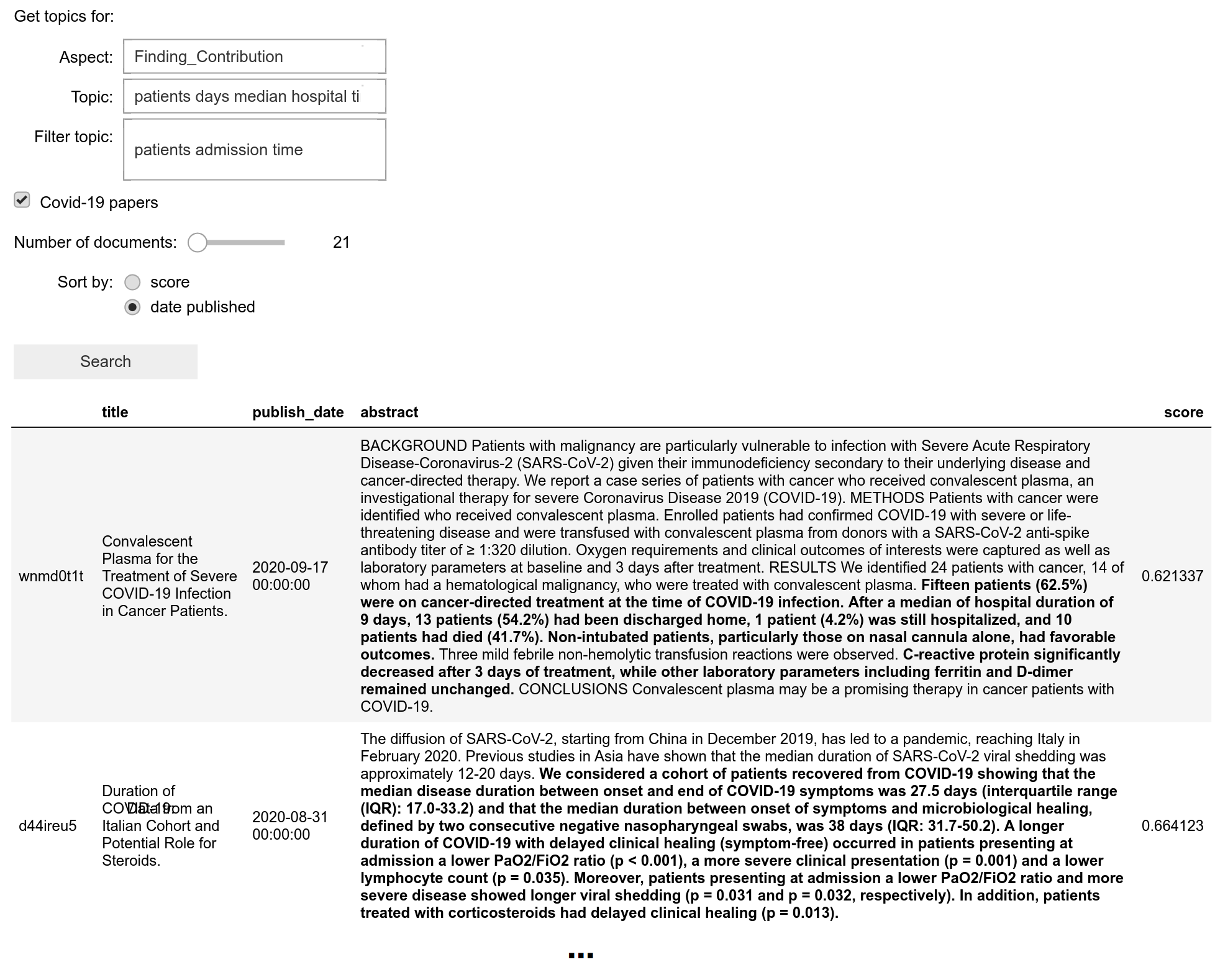}
  \captionof{figure}{List of COVID-19 abstracts whose finding/contribution addresses the topic ``patients, day, median, hospital, iqr, admission, range, died and years'', ordered by date of publication}
  \label{fig:patients}
\end{figure*}

\begin{figure*}
  \centering
  \includegraphics[width=.7\linewidth]{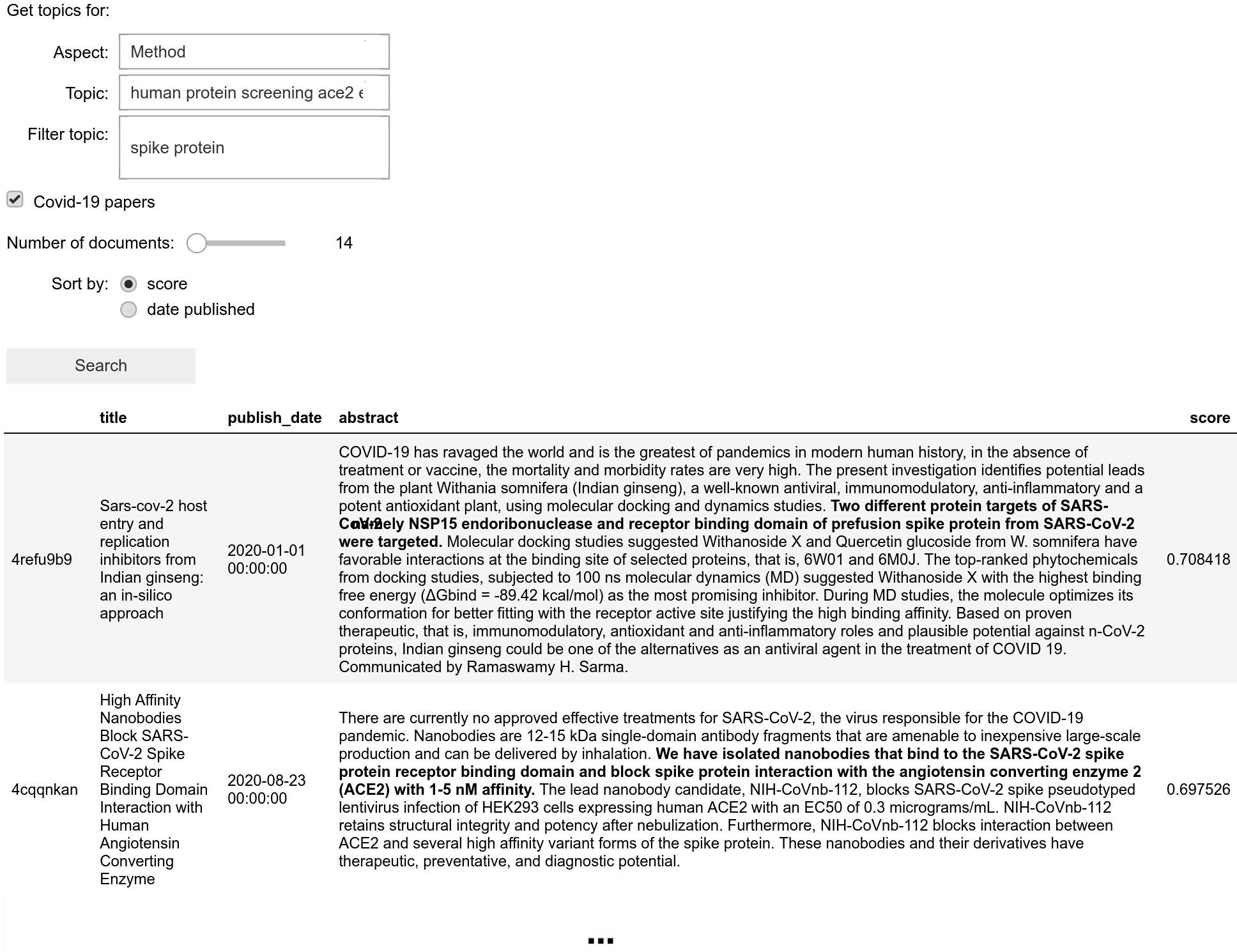}
  \captionof{figure}{List of COVID-19 abstracts whose finding/contribution addresses the topic ``human, protein, screening, ace2 expression, compounds, spike, active, proteins and screened'', ordered by score}
  \label{fig:spike}
\end{figure*}

\begin{figure*}
    \centering
    \includegraphics[width=.7\linewidth]{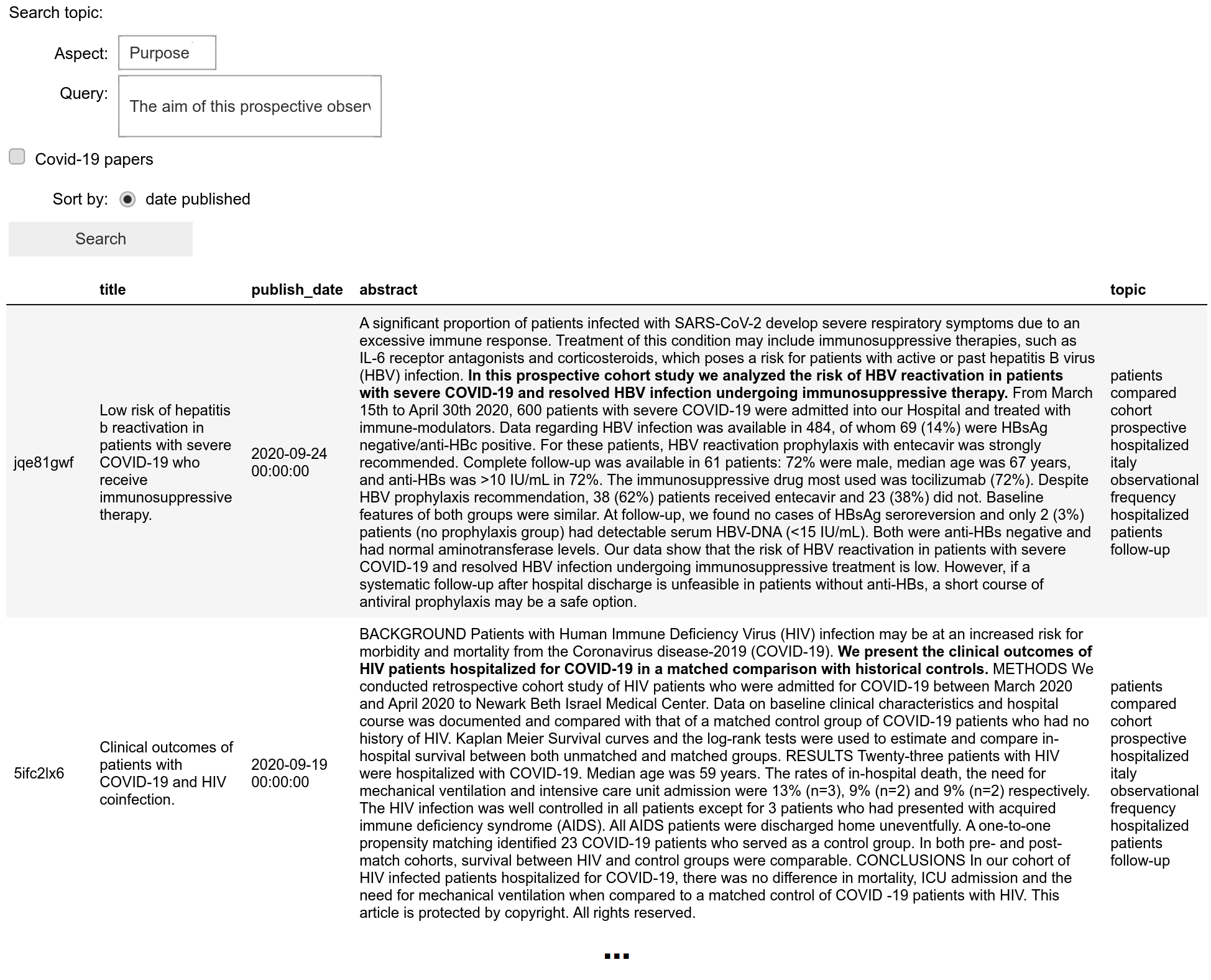}
    \captionof{figure}{List of recommended papers similar to query text}
    \label{fig:query}
\end{figure*}

\subsection{List papers in a topic}
The researcher can obtain the papers that best reflect a selected topic. Like the previous use case, filtering by research aspects and COVID-19 only papers is possible. Additionally, the researcher can filter topics that contain a query keyword. The score shown is the affinity of each paper to the topic in question.

The case study shown in Figure \ref{fig:patients} shows a search for the 19 most relevant documents about COVID-19 in a topic related to the finding/contribution research aspect. The researcher used the ``filter topic'' text box and entered the word ``patients admission time'', meaning only topics containing these words will be available for selection. 

In this case, the researcher opted to order the document list by publication date, meaning the most recent paper that addresses the topic is shown at the top. The earliest published paper does show a study about a treatment method that resulted in a reduction of the number of days the patients were hospitalized. This finding is arguably related to the selected topic, comprised of the words: ``patients, day, median, hospital, iqr, admission, range, died and years''.

Another example of this search is presented in Figure \ref{fig:spike}. The term ``spike protein'' is used as a filter. The researcher is interested in knowing all the studies done on this protein, which is being used in candidate vaccines \cite{Novavax2020NovavaxSite}. The topic comprised the words: ``human, protein, screening, ace2 expression, compounds, spike, active, proteins and screened'' was selected. The abstracts from the list all describe, in their methods, techniques like spike protein binding, expressing spike proteins and protein-protein interactions.

\subsection{Recommend reference papers based on a query text}
The researcher can search for papers that are the most similar to a query sentence given by them. As with previous use cases, the researcher can choose a specific aspect or the whole abstract for searching and narrow their search to only COVID-19 papers.

The researcher may seek prospective observational studies about COVID-19 and related coronavirus. He can provide a query text for that: ``The aim of this prospective observational cohort study...''. Figure \ref{fig:query} shows that this search could return the documents of their interest.

\subsection{Search papers based on a research question}
A list of predefined terms are given to the user to select from. This search returns papers where the term selected is present. As with previous use cases, the researcher can choose a specific aspect or the whole abstract for searching and narrow their search to only COVID-19 papers.

\begin{figure*}
  \centering
  \includegraphics[width=0.7\linewidth]{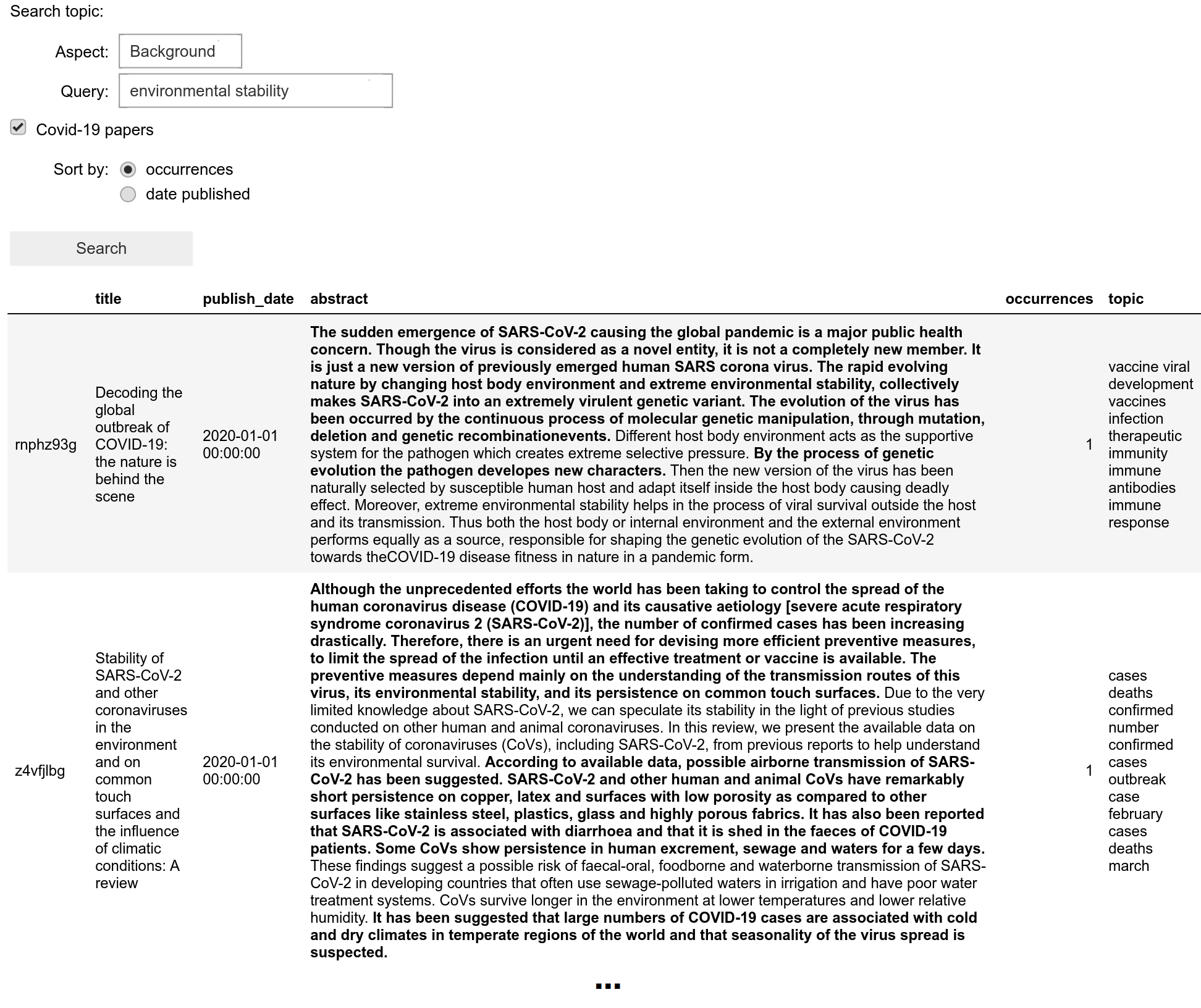}
  \captionof{figure}{List abstracts whose background addresses the topic of environmental stability of the virus}
  \label{fig:envir}
\end{figure*}

Figure \ref{fig:envir} outlines a search about the ``environmental stability'' of the virus.

\subsection{Plot documents in topic model space}

\begin{figure*}
\centering
\includegraphics[width=0.4\linewidth]{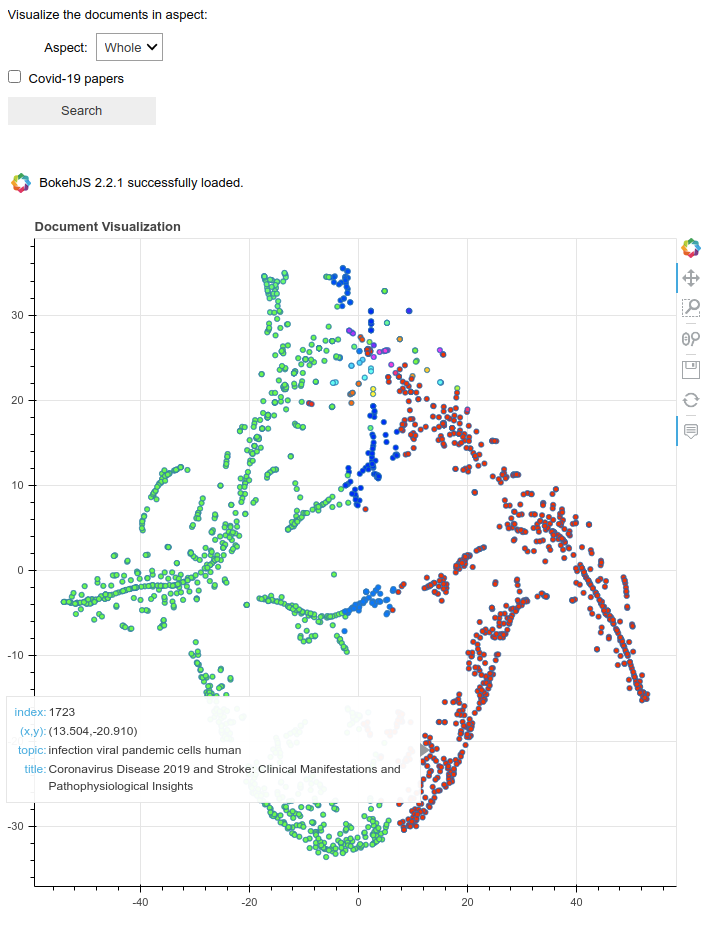}
\captionof{figure}{Interactive visualization of documents in research aspect}
\label{fig:viz}
\end{figure*}

The plot in Figure \ref{fig:viz} expresses each abstract as a point in 2D space with an assigned color representing a specific topic. The researcher can choose a specific aspect or the whole abstract for plotting and narrowing down their visualization to only COVID-19 papers. Titles and topics are visible when a point is hovered with the mouse cursor.

\subsection{Single abstract visualization}
\begin{figure*}
\centering
\includegraphics[width=.85\textwidth]{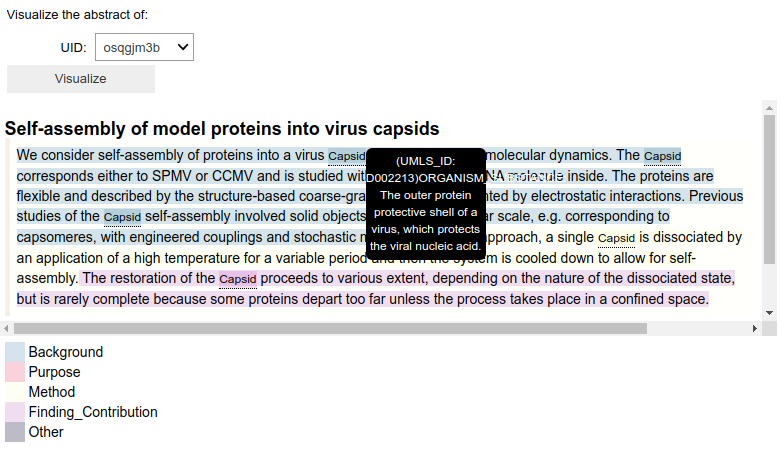}
\captionof{figure}{Individual abstract visualization}
\label{fig:single}
\end{figure*}

Figure \ref{fig:single} shows a case where the researcher looks at an individual abstract. He can see the individual role of each sentence (background, purpose, method, finding and other) based on its color. Hovering over an underlined entity shows the UMLS concept type, ID, and description of the corresponding entity.

\section{Discussion}

\subsection{Dataset choice}

CORD-19 is an open-access, curated, and constantly updated data set. Another important characteristic that it possesses is its broad set of topics \cite{Colavizza2020ACORD-19}. Given its consistency in providing good-quality papers on the topic, the data set fits the core purpose of this work, which is to filter information. To discern between abstracts about COVID-19 and abstracts about all coronavirus-related studies in CORD-19, we had to find those abstracts that contained the words ``COVID-19'' or ``COVID''. This approach can return false positive results simply because the authors can refer to one of these terms without them being the central theme of the study. This can be solved if we use a COVID-19 specific data set like LitCovid \cite{Chen2020KeepResearch}

It is important to note that CORD-19 does not include other data sources besides scientific articles. CORD-19 does not contain social media posts, forums, or articles. It also does not include preprint papers.

\subsection{LDA topic modeling}

Topic quality and count is very dependent on CODA-19 classifier's ability to recognize a particular research aspect. Research aspects with higher accuracy scores, such as findings and purpose, had more documents for training and led to more coherent topics.

A certain degree of arbitrariness was present in the parametrization step of the LDA topic modeling. The number of topics chosen for each research aspect, for example, was chosen using the default value of the \textit{ktrain} tool. The number of topics often depends on what the tool is trying to achieve. A greater number leads to less general topics and the splitting of similar documents to slightly more specific topics. A lower topic count yields more general groupings.

Methods that measure the coherence of the topic modeling to better determine the number of topics could be used, such as the one described by \cite{Mimno2011OptimizingModels} and \cite{Roder2015ExploringMeasures}. Visualization tools like LDAvis \cite{Sievert2014LDAvis:Topics} can be useful in manually analyzing topic coherence but require expert opinion. For the pruning parameters (min\_df and max\_df) the default value was used. Changing these two parameters to more conservative values does not decrease the quality of the model, but it can speed up the topic modeling process. We recognize that fine-tuning the parameters, such as the number of topics and iterations, could lead to better topic models and, thus a more accurate tool.

\subsection{Research aspect classification}

The performance of the CODA-19 SciBert pre-trained model for identifying research aspects has an average accuracy of 0.774, an overall satisfactory result. The model is slightly better at identifying findings and background.

The model has the potential to be improved. It is important to note that volunteers annotated the research aspects without prior knowledge of the subject matter. The agreement between annotators was considerable (Cohen's kappa of 0.677), but indicates the possibility of improvements in the model. The search functionality of the tool might become more reliable and accurate with future improvements to the model's accuracy.

Considering only abstract text for analysis and not the full text can lead to problems. There are instances where research aspects are left out from the abstract by the study's author. These missing research aspects could be extracted from the full text. We also noted that, when explicitly written, the abstract sections (BACKGROUND, OBJECTIVES, METHOD, DISCUSSION, and CONCLUSIONS) could be used with CODA-19's classifier to improve its accuracy.

\subsection{Named Entity Recognition and UMLS linking}

The NER approach used in the tool can recognize terms from various biomedical conceptual systems, describe them, and link them to UMLS. We realize that new terms are being created to express new findings in COVID-19 research. SciSpacy's NER might not recognize said terms. Because of this, more up-to-date solutions are important. Not only that, datasets focused on COVID-19 studies, such as CORD-NER \cite{Wang2020ComprehensiveSupervision}, achieved a higher F1 score compared to SciSpacy's annotation method, and show an improvement over broader NER solutions. Further development of the tool could leverage such resources.

\subsection{Documents visualization}

The graphical visualization allows researchers to explore the hidden relations identified by the topic model. Although less obvious to the end user than the other features in the tool, the graph can be used to discover unforeseen patterns relevant to that research aspect.

\subsection{Relevant document search}

It was thought from the start that modeling the topics based on research aspects would make the search of relevant documents more fine-grained and as a result, more accurate. While this statement might be true, it requires further investigation. We noticed that this strategy is heavily dependent on the topic model quality. Exploration of COVID-19 only papers yields more coherent results than searching all CORD-19 papers. This might be caused by the extremely broad topics addressed in coronavirus-related diseases.

The cases presented in the result section represent ideal search situations. We identified a couple of problems in our approach. For instance, the topics shown in the search functionality are represented as a collection of words. This collection can be confusing to the researcher and hard to interpret. We know that topic label generators \cite{Alokaili2020AutomaticLabelsb} can solve this problem by building a coherent sentence that explains the topic better. The K-Nearest Neighbor search based on topic probability distribution is also questionable. Other representations could be used to measure textual content similarity, such as word embeddings or sentence embeddings from BioSentVec \cite{Chen2018BioSentVec:Texts}.

The search with predefined terms from Kaggle, although simple, can return papers that address the terms chosen by the researcher by keyword matching. New research questions may arise, making it necessary to update this list.

\subsection{Single abstract visualization}

Highlighting the research aspects of the abstract speeds up the identification of the most important sections for the researcher, such as scientific finding, method etc. The description of the named entities simplifies the acquisition of knowledge about the a particular entity in the text. Linking the named entity to an UMLS entry allows the researcher to access the term directly in UMLS.

\section{Conclusion}

The worldwide crisis the new coronavirus has caused, in recent years, very negative effects to the world's health systems and areas such as economy and education. The disease is the subject of an increasing number of articles that describe new findings about the topic. The need for automated tools to support the researcher has been deemed important to tackle this increase in COVID-19 paper data. Such tools can apply machine learning algorithms to improve the search of COVID-19 documents and extract relevant information. The solution developed in this study was oriented towards the solution to this problem.

We can use several techniques to add value to such an application. The literature demonstrates that topic modeling, extraction of named entities, extraction of relations, detection of contradictory text, models for question answering, and other PLN algorithms have the potential to fill the gaps present in this type of solution. We demonstrated that combining different algorithms, libraries, and models can enhance the potential of information retrieval and extraction from articles about COVID-19.

\bibliography{template}  
\end{multicols}

\end{document}